# Light transmission behaviour as a function of the homogeneity in one dimensional photonic crystals

Michele Bellingeri[1], Francesco Scotognella*,[2]

[1]Dipartimento di Scienze Ambientali, Università di Parma, Parco Area delle Scienze, 33/A 43100 Parma, Italy

[2]Dipartimento di Fisica, Politecnico di Milano, piazza Leonardo da Vinci 32, 20133 Milano, Italy

* Corresponding author: francesco.scotognella@polimi.it

**Abstract**

The average light transmission of one-dimensional photonic media has been studied as a function of the medium homogeneity, quantified by the Shannon-Wiener index. We have found a decrease in the average light transmission by increasing the Shannon-Wiener index up to minimum (corresponding to *H'*=0.9375): from this point, the transmission increases following the Shannon-Wiener index. The behaviour has been confirmed for different pairs of materials forming the photonic structure. Nevertheless, we have observed that the trend slope is proportional to the refractive index ratio between the two materials ($n_{hi}/n_{low}$).

# 1. Introduction

The study of electromagnetic wave propagation in complex and structured photonic media is a highly relevant research field since it can improve the understanding of some general properties of transport phenomena. [1-3] Complex dielectric structures show variations of the refractive index on a length scale comparable to the wavelength of light. In structures with an ordered dielectric periodicity, namely photonic crystals, for a certain range of energies and certain wave vectors, light is not allowed to propagate through the medium [4-6]. Such behaviour is very similar to the one of electrons in a semiconductor material, where energy gaps arise owing to the periodic crystal potential in the atomic scale. Photonic crystals exist in nature or can be fabricated using a wide range of techniques, with the dielectric periodicity in one, two and three dimensions [7-9]. In the one-dimension case, simple and low-cost fabrication techniques can be used, as for instance spin coating or co-extrusion [7,10] Nowadays, these materials are extensively studied since they find application in several fields, including photonics for low threshold laser action, high bending angle waveguide, super-prism effect, sensors and optical switches. [11-16] The optical properties of photonic crystals, as for example the transmission of light, can be predicted by several mathematical methods [6,17-20]. Yet, these calculations can become very cumbersome as regards less homogeneous structures. Simple and not time consuming methods can be very useful for a better comprehension of the optical properties of such complicated systems. Recently, concepts and methods widely used in statistics have been successfully applied to explain light transport phenomena in Lévy glasses [21,22].

Herein, we have studied the light transmission properties of one dimensional photonic media, demonstrating a scaling law between the average transmission of light over a wide range of wavelengths and the distribution of the diffractive elements in the photonic lattice, i.e. the homogeneity grade of the structure being quantified by the Shannon index, commonly employed in statistics and information theory [23]. We have calculated light transmission in such structures by using a finite element method. In particular, we have shown that the light transmission decreases linearly by increasing the Shannon index, i.e. by increasing the homogeneity of the pillars distribution in the crystals. Interestingly, the result is inverse to what has been observed in two-dimensional photonic media [24,25].

# 2. Outline of the Method

In this study, we consider a one-dimensional photonic crystal made of alternated layers of two different materials [6]. The two layers are made of Titanium dioxide ($n_T$ = 2.45) and Silicon dioxide ($n_S$ = 1.46). In order to have a lattice constant $a$ = 200 nm, and in order to satisfy a geometrical setting $n_Z d_Z \sim n_S d_S$, the thickness of Titanium dioxide is $d_T$ = 75 nm, whereas the thickness of Silicon dioxide is $d_S$ = 225 nm [6]. In Figure 1, SiO$_2$ layers are represented as three layers with a thickness of 75 nm each: it is thus evident that the crystal unit cell can be divided in four equal layers, three of them made of SiO$_2$ and one of TiO$_2$.

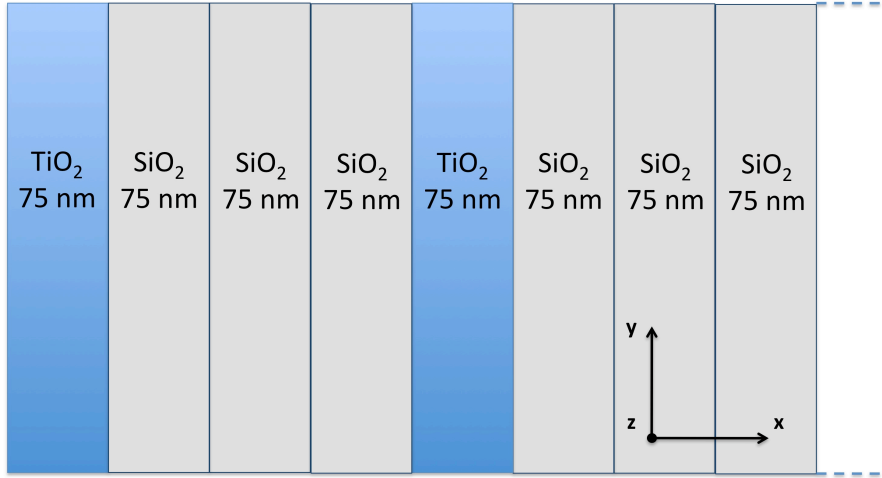

**Figure 1.** One-dimensional photonic crystal made of alternated Titanium dioxide and Silicon dioxide layers.

It is possible to correlate the distribution of the layers in the photonic structure to the Shannon–Wiener index [23]. The Shannon–Wiener H' index is a diversity index widely used in statistics and in information theory, and it is defined as

$$H' = -\sum_{j=1}^{s} p_j \log p_j \qquad (1)$$

where $p_j$ is the proportion of the $j$-fold species and $s$ is the number of the species. Dividing $H'$ by $log(s)$ we can normalize the index constraining it within the range (0,1). In previous works, we used the normalized Shannon index (i.e. $0 \leq H' \leq 1$) as a measurement of the homogeneity of two-dimensional media [24,25]. In those experiments we have distributed TiO$_2$ layers, i.e. the $j$-fold species, in the $s$ lattice cells. In the one-dimensional structure we can compute $H'$ by dividing the crystal length in a certain number of $s$ linear sub-units. Furthermore, we consider the Titanium dioxide layers as the $j$-fold species. The fraction of the layers belonging to each sub-unit represents the proportion $p_i$ in Equation (1). Ideally, $H'$ is the maximum when all sub-units contain the same number of layers. On the other hand, when all the layers are in one sole sub-unit, *H'* becomes the minimum: this structure is the most possible aggregated. As we said above, in this study we have made one-dimensional crystals with 64 layers, 16 are Titanium dioxide and 48 Silicon dioxide layers.

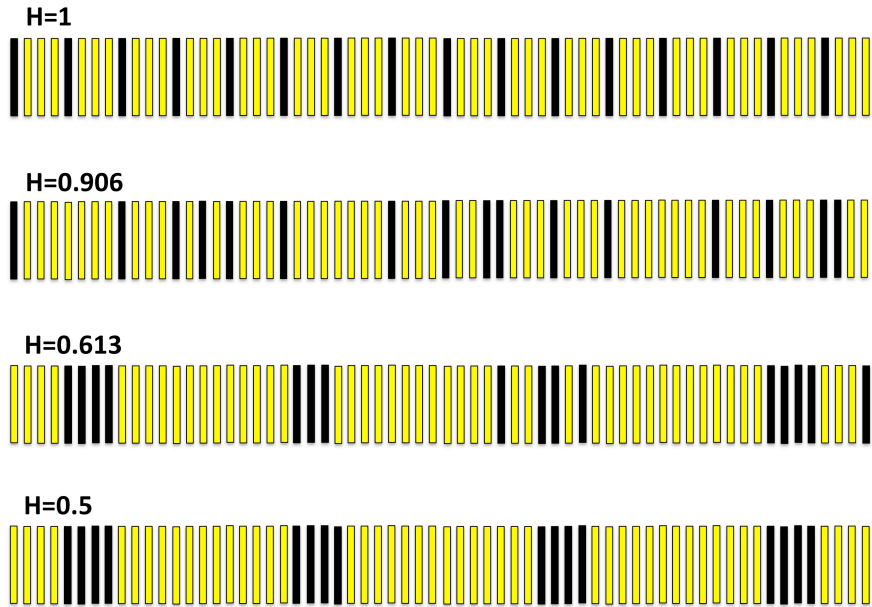

**Figure 2.** Scheme of one-dimensional photonic structures with different homogeneity.

Since we split the crystal in 16 sub-units, the most uniform crystal has a $TiO_2$ layer for each sub-unit. This homogeneous structure presents *H'* equal to 1. Conversely, since each sub unit can contain up to four pillars, the most aggregated linear configuration we can make is a crystal where four sub-units enclose four $TiO_2$. For the crystal topology selected in this study, the aforementioned configuration is the most non-homogeneous one, with *H'*=0.5. In this work, the crystal set is correlated to *H'* ranging in the interval (1,0.5). Figure 2 represents one-dimensional crystals with various grade of homogeneity.

For this study we have selected eleven different grades of homogeneity, i.e. *H'* = (1; 0.969; 0.9375; 0.906; 0.875; 0.83; 0.76; 0.66; 0.613; 0.55; 0.5). For each crystal (i.e. each Shannon index), five different realizations have been considered by allocating the $TiO_2$ layers in a random fashion within the belonging sub-unit. Thus, we have five different crystals and five different structures with the same global homogeneity and equal Shannon index. In other words, the number of layers in each sub-unit is the same in benchmark crystal and in its permutations.

For the calculation of the light transmission of the photonic structures through the finite element method, we assumed a TM-polarized field and we used the scalar equation for the transverse electric field component $E_Z$

$$\left(\partial_x^2 + \partial_y^2\right)E_Z + n^2 k_0^2 E_Z = 0 \qquad (2)$$

where $n$ is the refractive index distribution and $k_0$ is the free space wave number [6,26]. As regards the input field, a plane wave with wave vector $k$ directed along the $x$-axis has been assumed. Scattering boundary conditions in the $y$ direction have been used. For a comparison with the $TiO_2$-$SiO_2$ photonic media, we have performed the same light transmission simulations for other two couples of materials. A photonic medium has been made of Zinc Oxide ($n_Z$=2) and $SiO_2$, while the other has been made of ZnO and Poly(hexafluoropropylene oxide) (PHFPO, $n_P$=1.301), a polymer with a low refractive index. Therefore, we have analysed photonic media with three different refractive index ratios (1.678 for $TiO_2$/$SiO_2$, 1.537 for ZnO/PhFPO, 1.37 for ZnO/$SiO_2$).

## 3. Results and discussion

Figure 3 shows the transmission spectra of one-dimensional photonic media, made of alternated layers of $TiO_2$ and $SiO_2$, with three different homogeneities, i.e. $H'$=1 (black line), $H'$=0.9375 (red line) and $H'$=0.5 (green line). We selected these three homogeneities since they display the most significant dissimilarities in their transmission spectra between the grades of homogeneity chosen in this study. Furthermore, they correspond to the extremes in the trend represented in Figure 4 (see below). For $H'$=1, i.e. for the ideal photonic crystal, we have noticed a wide photonic band gap at around 1020 nm, that is in good agreement with the Bragg Snell law, i.e. $\lambda_{Bragg} = n_{eff}\Lambda$, where $\lambda_{Bragg}$ is the centre wavelength of the stop band, $n_{eff}$ is the effective refractive index of the lattice and $\Lambda$ the spatial period (in this case, $\Lambda = a = 300$ nm). On the contrary, the photonic medium with $H'$=0.9375 has more several transmission features that are not present for the ideal photonic crystal. There is still a photonic band gap, even if blue shifted, as well as two transmission peaks around 900 nm. Yet, a new feature at 750 nm and an other one at 1250 nm occur. Finally, the photonic structure with homogeneity corresponding to $H'$=0.5 almost show two weak bands at about 800 nm and 1000 nm.

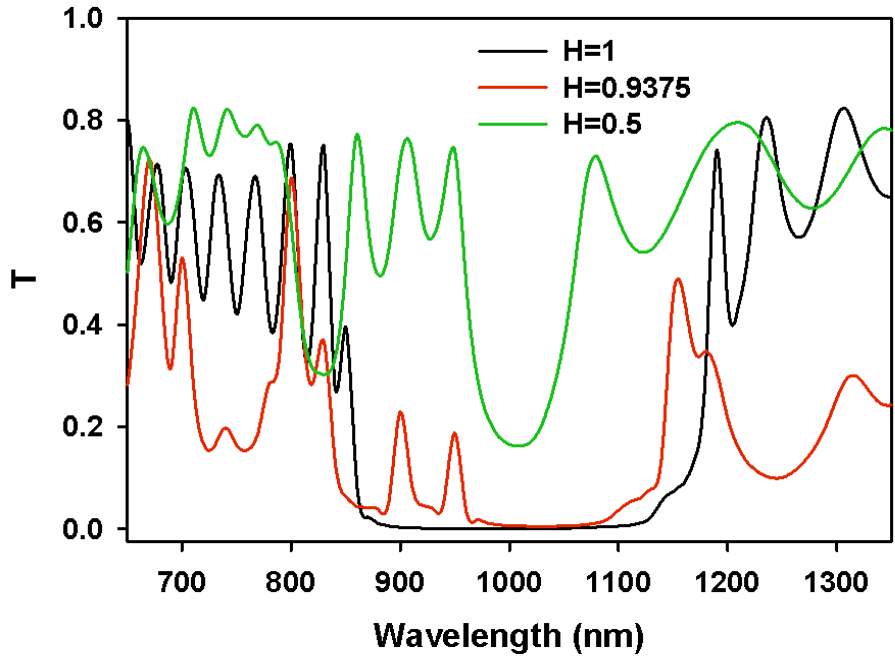

**Figure 3.** Transmission spectra of two one-dimensional photonic structures with different homogeneity.

Figure 4a represents the average light transmission in the range 650-1350 nm as a function of the Shannon index for $TiO_2/SiO_2$ photonic media, normalized to the average light transmission of the ideal photonic crystal. The five points for each Shannon index value (except for *H'*=1, for which it is not possible to permute the structure for topological reasons) correspond to the five different realizations, i.e. to the five crystals of equal homogeneity. The black line connects the mean values of the light transmission. The fact that the behaviour is consistently dissimilar with respect to the results obtained for two-dimensional photonic media deserves consideration. A linear increase of the average light transmission as a function of the Shannon index has been observed in the two-dimensional case [23]. Yet, in the one-dimensional case, we have observed a decrease of the light transmission in the range *H'*=0.5÷0.9375 as a function of the Shannon index. Subsequently, we have observed an increase of the light transmission in the range *H'*=0.9375÷1. In this trend, the value of *H'*=0.9375 is a minimum.

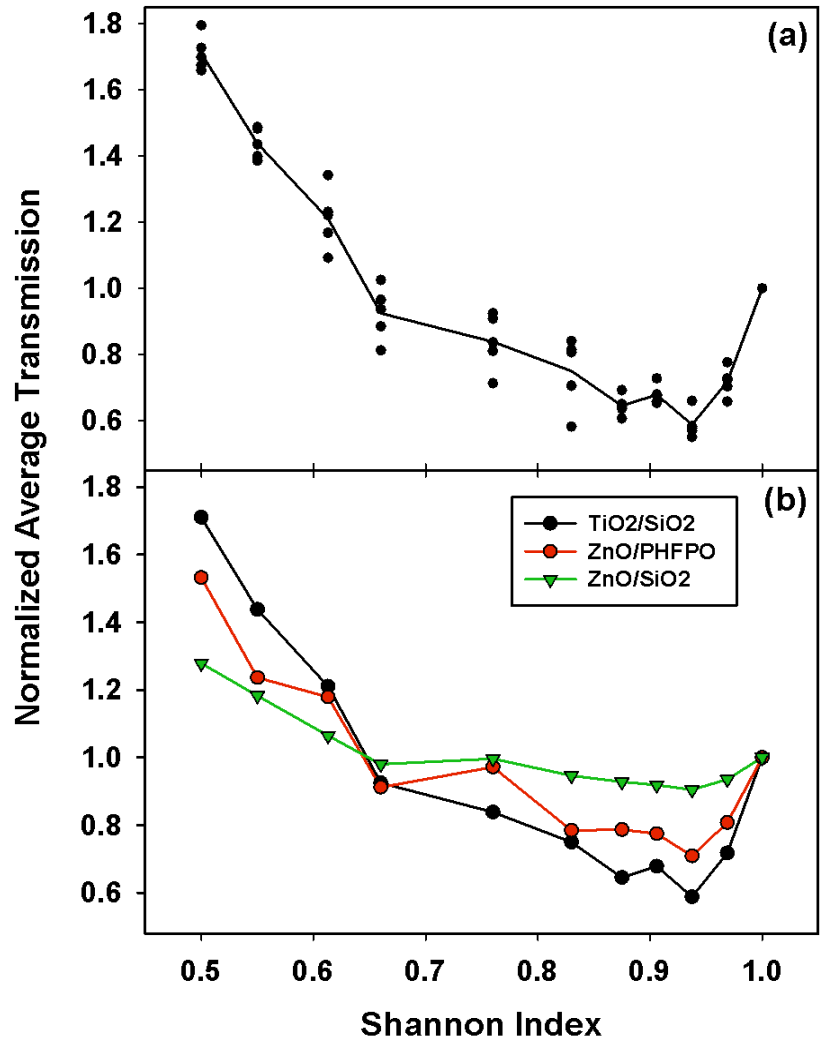

**Figure 4.** Normalized average transmission as a function of the Shannon index.

For a better understanding of the results obtained with this *in silico* experiment, we have analysed the trend of the average light transmission as a function of the Shannon index for different pairs of materials, which are ZnO/SiO$_2$ and ZnO/PHFPO. It is noteworthy that, for the three different pairs of materials, the trend is invariant, as shown in Figure 4b, with the same minimum at *H'*=0.9375. Thus, the average light transmission as a function of *H'* is independent on the refractive index ratio in the one-dimensional photonic crystal, which means that is independent on the materials chosen to make the photonic medium. Moreover, we have observed that the decaying slope of such trends increases with the refractive index ratio by normalizing the three trends to the average light transmission of the ideal photonic crystal. In other words, the average light transmission function for TiO$_2$/SiO$_2$ structure shows a sharper derivative with respect to the one for ZnO/SiO$_2$ structure.

## Conclusions

In this study we have engineered one-dimensional photonic structures with different grade of homogeneity. Such homogeneity is quantified by the Shannon index, which is commonly used in statistics and information theory. By means of a finite element method, we have simulated the optical properties of the engineered one-dimensional structure. Moreover, we have observed that the average light transmission as a function of the Shannon index shows a trend that is dissimilar from the one already reported for two-dimensional structures [24,25]. Average light transmission decreases by increasing the Shannon index up to $H'$=0.9375, while it is growing with a Shannon index in the range (0.9375,1). Moreover, this trend behaviour is independent of pairs of materials chosen to build the photonic structure, even if the trend slope raises by increasing the refractive index ratio.

## Acknowledgements

The authors acknowledge Emanuela Tenca, Ajay R. Srimath Kandada, Prof. Guglielmo Lanzani and Prof. Stefano Longhi for helpful discussions.